\documentclass[fleqn,usenatbib]{mnras}
\usepackage{newtxtext,newtxmath}

\usepackage[T1]{fontenc}

\usepackage{ae,aecompl}
\usepackage{comment}

\usepackage{microtype}
\usepackage{amsmath}
\usepackage{graphicx}
\usepackage{lscape}

\usepackage{color}

\usepackage{url}

\graphicspath{{pics/}}
\newcommand {\bc}{\begin {center}}
\newcommand {\ec}{\end {center}}
\newcommand {\be}{\begin {equation}}
\newcommand {\ee}{\end {equation}}
\newcommand {\beq}{\begin {eqnarray}}
\newcommand {\eeq}{\end {eqnarray}}

\definecolor{mypink1}{rgb}{0.858, 0.188, 0.478}

\def\flux{erg~s$^{-1}$~cm$^{-2}$}
\def\lum{erg~s$^{-1}$}

\def\xte{XTE\,J1859$+$083}

\title[X-ray pulsar \xte]
{On the nature of the X-ray pulsar \xte\ and its broadband properties}

\author[A.~Salganik et al.]
{Alexander~Salganik,$^{1,2}$\thanks{E-mail: alsalganik@gmail.com} 
Sergey~S.~Tsygankov,$^{3,2}$
Anlaug~A.~Djupvik,$^{4,5}$
Dmitri I. Karasev,$^{2}$
\newauthor
Alexander~A.~Lutovinov,$^{2}$
David~A.~H.~Buckley,$^{6,7,8}$
Mariusz~Gromadzki,$^{9}$
and Juri Poutanen$^{2,3,10}$
\\
$^1$Department of Astronomy, Saint Petersburg State University, Saint-Petersburg 198504, Russia\\
$^2$Space Research Institute of the Russian Academy of Sciences, Profsoyuznaya Str. 84/32, Moscow 117997, Russia\\
$^3$Department of Physics and Astronomy,  FI-20014 University of Turku, Finland\\
$^4$Nordic Optical Telescope, Apartado 474, 38700 Santa Cruz de La Palma, Santa Cruz de Tenerife, Spain
\\
$^5$Department of Physics and Astronomy, Aarhus University, NyMunkegade 120, DK-8000 Aarhus C, Denmark
\\
$^{6}$South African Astronomical Observatory, PO Box 9, Observatory Road, Observatory 7935, Cape Town, South Africa\\
$^{7}$Department of Astronomy, University of Cape Town, Private Bag X3, Rondebosch 7701, South Africa\\
$^{8}$Department of Physics, University of the Free State, PO Box 339, Bloemfontein 9300, South Africa\\
$^{9}$Astronomical Observatory, University of Warsaw, Al. Ujazdowskie 4, 00-478 Warszawa, Poland\\ 
$^{10}$Nordita, KTH Royal Institute of Technology and Stockholm University, Roslagstullsbacken 23, SE-10691 Stockholm, Sweden
}
\date{Accepted 2021 November 15. Received 2021 November 15; in original form 2021 October 18}

\pubyear{2021}

\begin{document}
\label{firstpage}
\pagerange{\pageref{firstpage}--\pageref{lastpage}}
\maketitle

\begin{abstract}
This work is devoted to the study of the broadband 0.8--79~keV spectral and timing properties of the poorly studied X-ray pulsar \xte\ during its 2015 outburst based on the data from the {\it NuSTAR} and {\it Swift} observatories. 
We show that the source pulse profile has complex shape that depends on the energy band. Pulse fraction of \xte\ has constant value around 35\% in the broad energy band, this behaviour is atypical for X-ray pulsars. At the same time its energy spectrum is typical of this class of objects and  has a power-law shape with an exponential cutoff at high energies. No cyclotron absorption line was discovered in the source spectrum. On the basis of indirect method and the absence of a cyclotron line, an estimation was made for the magnetic field strength as less than $5\times10^{11}$~G or belonging to the interval from $5\times10^{12}$ to $2.0^{+0.9}_{-1.2}\times10^{13}$~G. 
Data from the NOT and SALT telescopes as well as optical and IR sky surveys allowed us also to study the nature of its optical companion. 
We have proposed and studied new possible candidates for the optical companion of \xte\ and the most likely candidate was identified. The results of the optical and IR photometry and spectroscopy of these possible companions  showed that the system is a Be X-ray binary, showing Br$\gamma$, \ion{He}{i} and strong H$\alpha$  spectral lines. 
\end{abstract}


\section{Introduction}
\label{intro}
The transient X-ray pulsar (XRP) \xte\ with a pulsation period of 9.8~s was discovered in August 1999 using the Proportional Counter Array (PCA) on board the {\it Rossi X-ray Timing Explorer} ({\it RXTE}) \citep{Marshall1999} and localized with coordinates RA $=18^{\rm h}59\fm1$, Dec. = $8{\degr}15\arcmin$ with error radius of $2'$ at a 90\% confidence level.
The transient nature of the source was confirmed by deep observation performed in 2007 with the XRT telescope on board of the {\it Neil Gehrels Swift} observatory, which made it possible to set a $3\sigma$ upper limit on the 0.3--10~keV flux from the source in the quiescent state of $5 \times10^{-14}$~\flux\  \citep{Romano2007}. 
Based on the data from a long-term monitoring of the pulsar by the  All-Sky Monitor (ASM) aboard the {\it RXTE} observatory, \citet{Corbet2009} estimated the orbital period of the system $P_{\rm orb} = 60.65 \pm 0.08$~d, consistent with the Be-nature of the potential optical companion. The authors also noticed a non-monotonic change in the pulsation frequency, most likely associated with the motion of the neutron star (NS) in the binary system. 
More accurate localization of XTE\,J1859+083 with coordinates RA = $18^{\rm h}59^{\rm m}2$, Dec = $8{\degr} 14\arcmin$ with error radius of $1\arcmin$ (at 90\% confidence level) was obtained using the {\it BeppoSAX} observatory data \citep{Corbet2009}.

In February 2015, the all-sky monitor MAXI detected a new X-ray outburst from \xte\ \citep{Negoro2015}. The outburst was confirmed by observations with other instruments: {\it Swift}/BAT \citep{Krimm2015}, {\it Fermi}/GBM \citep{Finger2015}, {\it Swift}/XRT \citep{LiKong2015}, {\it INTEGRAL} \citep{Malyshev2015}. Based on the data from the {\it Swift}/XRT, \citet{LiKong2015} obtained an accurate localisation of the source at RA = $18^{\rm h} 59^{\rm m}01\fs57$, Dec. = $08\degr14\arcmin44\farcs2$ with the $1\farcs9$ error radius at a 90\% confidence level. This allowed the authors to propose the star USNO-B1.0 0982-0467424 (2MASS 18590163+0814444) as a possible optical companion in the system.

Applying the torque model to the spin period measurements made with  {\it Fermi}/GBM  at the beginning of 2015, \citet{Kuehnel2016} obtained an estimate for the orbital period $P_{\rm orb}$ in the system of about 38 days, as well as estimates of the others orbital parameters. At the same time the residuals of the observational data from the model demonstrate an additional periodicity of 65 days, which is close to the value of the orbital period obtained by \citet{Corbet2009}. \citet{Kuehnel2016} suggests that this may be a so-called "superorbital" period of the system. Similar estimates for the parameters of the binary system were independently obtained by Mark H. Finger within the framework of the {\it Fermi} Gamma-ray Burst Monitor (GBM) Accreting Pulsars Program (GAPP)\footnote{\url{http://gammaray.nsstc.nasa.gov/gbm/science/pulsars/lightcurves/xtej1859.html}}.

In this paper, we present results of the first detailed study of the properties of \xte\ in a wide energy range of 0.8--79~keV, carried out using data from the {\it Swift} and {\it NuSTAR} observatories, that allowed us to estimate some properties of the NS in the system. In addition, based on the infrared observations with the Nordic Optical Telescope (NOT) and the optical observations with the Southern African Large Telescope (SALT) we analysed properties of the potential companion star and made conclusions about the type of the system.   

\section{Data analysis}
\label{sec:data}

\begin{figure}
    \centering
    \includegraphics[width=0.95\columnwidth]{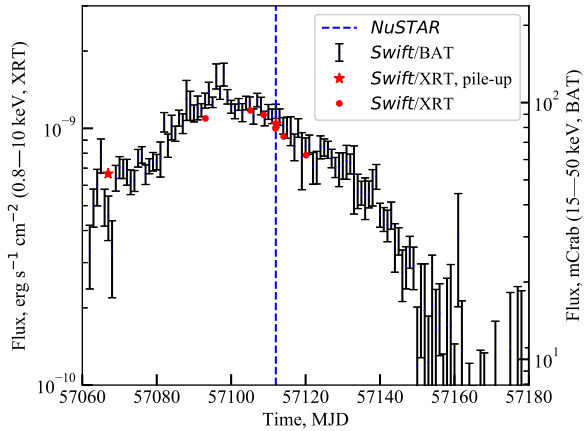}
 	\caption[The LOF caption]{Light curve of \xte\ according to the {\it Swift}/BAT telescope monitoring in the 15--50~keV energy range. The moment of observation of the source with {\it NuSTAR} is shown by vertical dashed line. {\it Swift}/XRT observations are superimposed for comparison.}
	\label{fig:lightcurve}
\end{figure}

\xte\ was observed during the 2015 outburst with several X-ray instruments. For our X-ray analysis we used simultaneous observations with the {\it NuSTAR} (ObsID 90001010002) and {\it Swift}/XRT (ObsID 00081447001) telescopes done on 2015 March 31 at the initial stage of the outburst decay. 

\subsection{{\it NuSTAR} observatory}

The \textit{NuSTAR} observatory includes two co-aligned X-ray telescopes focusing X-ray photons onto two Focal Plane Modules A and B (FPMA and FPMB) \citep{Harrison2013}. \textit{NuSTAR} energy range is 3--79~keV, in the context of the current work the most important \textit{NuSTAR} capability is its high sensitivity in the hard X-rays. 

The effective exposure time of the {\it NuSTAR} observation is 20\,ks.
The data were processed using the {\sc heasoft} package version 6.28 and the {\sc caldb} 4.9.0 calibration files package in accordance with the data analysis manual.\footnote{\url{https://heasarc.gsfc.nasa.gov/docs/nustar/analysis/nustar_swguide.pdf}}
The data for the analysis of the source were extracted from a region with a radius of 40\arcsec, the radius of the background region was 114\arcsec. The extraction regions were selected to maximize the signal-to-noise ratio at high energies. Energy channels were binned in accordance with the optimal binning algorithm proposed by \citet{Kaastra2016} using the {\sc ftgrouppha} utility from the {\sc ftools} package. The counts from the two telescopes were summed after the background subtraction and barycentric correction was done using the {\sc barycorr} utility.

\subsection{{\it Swift} observatory}
Observation of the {\it Swift}/XRT  (ObsID 00081447001) was performed in two modes simultaneously with \textit{NuSTAR}: Photon Counting mode (PC) with the exposure of 362\,s and Windowed Timing mode (WT) with the exposure of 572\,s. Since the source was bright, for the spectral analysis we used only data obtained in the WT mode to exclude the possible influence of the pile-up effect.
To follow the outburst evolution at soft X-ray energies we also used a series of the {\it Swift}/XRT observations (ObsID 00037043005, 00037043006, 00037043009, 00037043010, 00037043011, 00081447001, 00037043012, 00037043014) done in the WT mode  (Fig.~\ref{fig:lightcurve}). \textit{Swift}/BAT light curve\footnote{ \url{https://swift.gsfc.nasa.gov/results/transients/weak/XTEJ1859p083/}} is also plotted for comparison. The \textit{Swift}/XRT light curve was obtained by fitting the spectra with simple absorbed power-law model.

The {\it Swift}/XRT spectra were produced using the online service provided by the UK {\it Swift} Science Data Center \citep{Evans2009}.\footnote{\url{https://www.swift.ac.uk/user\_objects/}} The spectral channels were grouped in such a way that each channel had at least 1 count.
The resulting broadband spectrum of the source from all instruments was approximated with several continuum models using W-statistics\footnote{\url{https://heasarc.gsfc.nasa.gov/xanadu/xspec/manual/XSappendixStatistics.html}} \citep{Wachter1979} in the {\sc xspec} 12.11.1 package \citep{Arnaud1996}. All errors are given at the $1\sigma$ confidence level if not specified otherwise.

\subsection{Optical and IR sky surveys}

The magnitudes of stars in optical and near-IR ranges presented in the paper are taken from the public catalogs of sky surveys Pan-STARRS,\footnote{\url{https://panstarrs.stsci.edu}} and UKIDSS ESO.\footnote{\url{http://wsa.roe.ac.uk/}} 
We also use IPHAS sky survey data. Magnitudes of probable counterparts in H$\alpha$ and other filters of this survey, were determined through an additional photometric analysis of IPHAS image data using the PSF-photometry ({\sc daophot ii}). To convert the obtained instrumental magnitudes to the real ones, we match our results (for all stars of the field) with a standard IPHAS DR2 catalog\footnote{\url{https://cdsarc.cds.unistra.fr/viz-bin/cat/II/321}} and estimate the conversion factor between instrumental and real/observed magnitudes.
A search for an optical/infrared companion of the source based on the data from the Pan-STARRS, IPHAS and UKIDSS sky surveys revealed two potential counterparts (the Northern and the Southern, see Sect.~\ref{sec:iden}). 
Then, we determined magnitudes of the proposed counterparts in the H$\alpha$, $r$, and $i$ filters. The coordinates and distances to putative companion stars, among others, were obtained by \citet{Bailer2021} (see catalog \url{https://cdsarc.unistra.fr/viz-bin/cat/I/352}) based on \textit{Gaia}\footnote{\url{https://sci.esa.int/web/gaia}} Early Data Release 3 (EDR3).

\subsection{Nordic Optical Telescope}

The near-infrared Camera and Spectrograph, NOTCam \citep{Abbott2000}, with its Hawaii-1 HgCdTe detector was used at the 2.56m NOT \citep{Djupvik2010} to obtain K-band spectra on 2021 June 20 of both candidates inside the error circle of {\it Swift}/XRT. The instrument setup used was the WF-camera (0\farcs234/pix), the 128 micron wide slit (0\farcs6), Grism \#1, with the MKO K-band filter (NOT \#208) used as an order sorter. Grism \#1 has a dispersion of 4.1\AA \ per pixel in the K-band, giving a resolving power of $\lambda/\Delta \lambda$ = 2100 for our setup. 

The night was photometric and the seeing measured in the acquisition images had a FWHM = 0\farcs7. The rotator was oriented to include the two targets, separated by 1\farcs8, in the slit. The average airmass was 1.1. The spectra were obtained in an ABABAB dithering mode, exposing 300~s per individual spectrum using the ramp-sampling mode with 10 non-destructive readouts every 30~s. This cycle was repeated 2.5 times to provide 15 individual spectra. Arc and halogen lamps were observed while still pointing to the target, and the nearby A0~V star HD189920 was observed as a telluric standard just before the target. 

The data were reduced with own scripts within the {\sc iraf} package.\footnote{\url{https://iraf.noao.edu/}}
The individual exposures were corrected for hot pixels, using darks obtained with the same integration time to make hot pixel masks, flat-field corrected using the halogen flats, and sky-subtracted using the dithered neighbouring frame. The individual 1D spectra were optimally extracted and wavelength calibrated, after which they were combined to final spectra. The telluric standard was observed and reduced in a similar manner and used to correct the science spectra for the features produced by the Earth's atmosphere, removing first its Br$\gamma$ absorption line. 

For an approximate flux calibration of the spectra, we used a sample of 10 high quality 2MASS stars in the acquisition image in order to derive the magnitudes of the two targets to be 13.35 and 13.61 mag, respectively, for the Northern (component 1) and the Southern (component 2). We estimate the uncertainty to be 0.1 mag based on the 0.06 mag scatter in the offset between NOTCam and 2MASS magnitudes of the calibration stars and the slightly different K-band filters. After having divided the spectra by the A0~V standard, we multiply each target spectrum by its properly flux-scaled Vega continuum, which corrects the slope and provides an approximate flux calibration.

\subsection{Southern African Large Telescope spectroscopy}

Spectroscopy of the two optical counterpart candidates was undertaken with SALT \citep{Buckley2006SPIE.6267E..0ZB} on 2021 August 28. Two consecutive 1200 s exposures were obtained, beginning at 18:08:25 UTC, with the Robert Stobie Spectrograph \citep[RSS;][]{Burgh2003SPIE.4841.1463B} which used the PG900 VPH grating, covering the region 3920--6990~\AA\AA{} at a mean resolution of 5.7~\AA{} with a 1\farcs5 slit width.

The spectra were initially reduced using the {\sc pysalt} package
\citep{Crawford2010SPIE.7737E..25C},\footnote{\url{https://astronomers.salt.ac.za/software/pysalt-documentation/}} which undertakes bias, gain and amplifier cross-talk corrections, mosaics the three CCDs and applies cosmetic corrections. The spectral extraction, wavelength calibration and background subtraction were all undertaken using standard {\sc iraf} routines, as was the relative flux calibration. Due to the mediocre seeing of 2\arcsec, it was not possible to extract the spectra of the two stars separately, therefore the derived spectrum is for both stars combined.

\section{Results}

\subsection{Timing analysis}

Flux pulsations from \xte\ were searched in the {\it NuSTAR} data in the full energy range using the {\sc efsearch} tool from the {\sc heasoft} package. As a result, the pulsation period was determined $P_{\rm spin} = 9.79156 \pm 0.00001$ s.
The uncertainty for the period was estimated by simulating a large number of light curves obtained by varying the count rate from the source within the statistical error, followed by searching for the period in the simulated light curve. For a more detailed description of the method, see \citet{Boldin2013}.

\begin{figure}
\centering
\includegraphics[width=0.95\columnwidth]{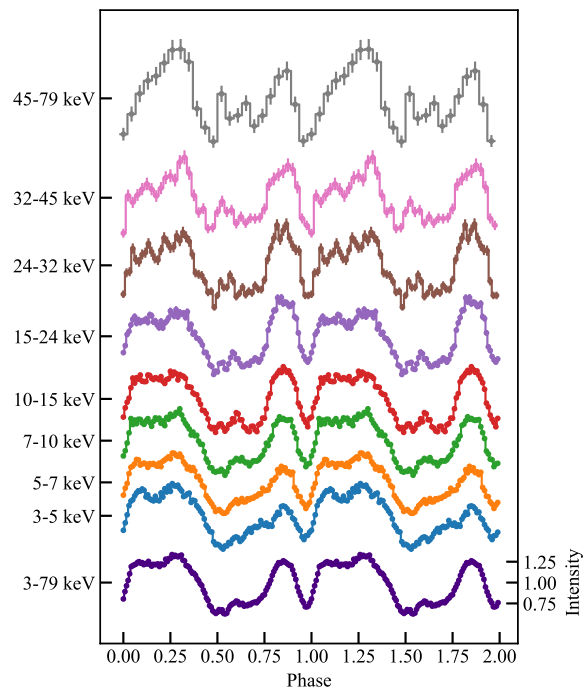}
\caption{Normalized pulse profiles of \xte\  in different energy bands according to the {\it NuSTAR} data.}
	\label{fig:profile}
\end{figure}

High count statistics allowed us to study the dependence of the source pulse profile on photon energy. In Fig.~\ref{fig:profile} one can find energy-resolved pulse profiles of \xte\ normalized by the average intensity in a given energy band.
We see that at soft energies the pulse profile in the first approximation can be described by two broad peaks with maxima at phases 0.1--0.3 and 0.8--0.9. At higher energies, a finer structure of the profile begins to appear with an increase of the relative contribution of the intermediate peak at phases 0.5--0.6. At the highest energies (above 40~keV), the contribution of the left wing of the first peak is significantly weakened. 

\begin{figure}
\centering
\includegraphics[width=0.95\columnwidth]{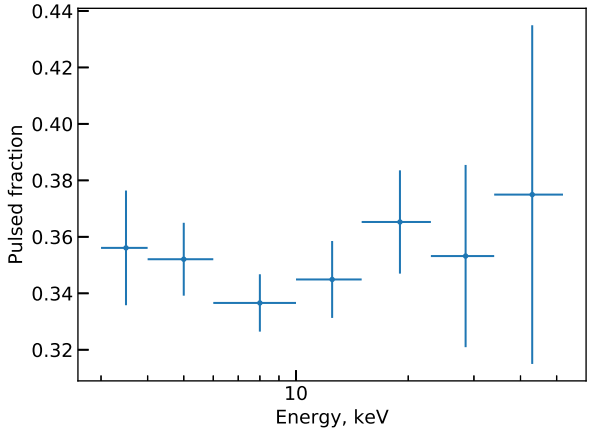}
\caption{Dependence of the pulsed fraction in \xte\ on the energy obtained from the {\it NuSTAR} data.}
\label{fig:pulsed}
\end{figure}

The energy-resolved count rate $C$ of the pulsar was also used to study the dependence of the pulsed fraction, defined as $({\max C - \min C})/({\max C + \min C})$, 
on energy, as shown in Fig.~\ref{fig:pulsed}. 
The pulsed fraction in each band was calculated using 15 phase bins in the pulse profile. We see that the pulsed fraction is practically independent of the energy, staying at around 35\%. Such behavior is atypical for most of the studied X-ray pulsars, where a significant increase of the pulsed fraction with the photon energy is usually observed \citep[see, e.g.,][]{LutovinovTsygankov2009}.

\begin{table*}
\caption{Best-fit spectral parameters of \xte\ for different continuum models.}
\begin{center}
\begin{tabular}{lccc} 
 \hline
 Parameter  & {\sc gabs $\times$ po $\times$ highecut} & {\sc comptt} 
 & {\sc cutoffpl} 
 \\
  \hline
 ${\rm const}_{\rm FPMA}$ & 1.000 (frozen)  &  1.000 (frozen) &  1.000 (frozen)  \\ 
 ${\rm const}_{\rm FPMB}$ & $1.042 \pm 0.003$ &  $1.042 \pm 0.003$ &  $1.041 \pm 0.003$ \\ 
 ${\rm const}_{\rm XRT}$ &  $1.005 \pm 0.018$  &  $1.036 \pm 0.018$ &  $0.985 \pm 0.018$  \\
 $N_{\rm H}$, $10^{22}$~cm$^{-2}$ & $3.1\pm0.1$ 
 & $3.9\pm0.1$ & $2.2\pm0.1$ \\
 $E_{\rm cut}$,~keV & $18.9\pm 0.3$ & &  \\
 $E_{\rm fold}$,~keV & $27.0\pm0.5$ & & $24.9\pm0.3$ \\ 
 $\Gamma$ & $1.40\pm0.01$ &  & $1.04\pm0.01$ \\ 
 $T_0$,~keV & & $0.22\pm0.08$ & \\ 
 $T$,~keV & & $8.41\pm0.05$ &   \\ 
 $\tau$ & & $4.43\pm0.02$ &  \\
 geometry $\beta$ & & 1.0 (frozen) &  \\
 ${\rm Norm}_{\rm continuum}$, ph~keV$^{-1}$~s$^{-1}$~cm$^{-2}$  & $0.123\pm0.001$ &  $0.054 \pm 0.011$ & $0.083\pm0.001$   \\
 
 $E_{\rm gabs}$,~keV & $18.9$ (=$E_{\rm cut}$) & & \\ 
 $\sigma_{\rm gabs}$,~keV & $1.9$ (=$0.1 \times E_{\rm cut}$) & & \\
 $\tau_{\rm gabs}$ & $0.06\pm0.02$ & & \\
 $E_{\rm gauss}$,~keV & $6.43 \pm 0.04$ &  $6.47 \pm 0.04$
  & $6.41 \pm 0.05$ \\
 $\sigma_{\rm gauss}$,~keV & $0.15 \pm 0.11$ &  $0.15$ (frozen) 
 & $0.15$ (frozen) \\
 ${\rm Norm}_{\rm gauss}$, ph~s$^{-1}$~cm$^{-2}$  & $(3.2 \pm 0.6 )\times 10^{-4}$& $(3.0 \pm 0.4 )\times 10^{-4}$ & $(1.8 \pm 0.4 )\times 10^{-4}$ \\
 Flux$_{0.8-79 \rm keV}$, \flux & $(2.65\pm0.01)\times10^{-9}$ & $(2.64\pm0.01)\times10^{-9}$ & $(2.64\pm0.01)\times10^{-9}$ \\
  W-statistic/d.o.f. & 1161/1129 & 1572/1131 & 1785/1132 \\
 \hline
\end{tabular}
\label{table:spec_params}
\end{center}
\end{table*}

\subsection{Spectral analysis}

\begin{figure}
\centering
\includegraphics[width=0.95\columnwidth]{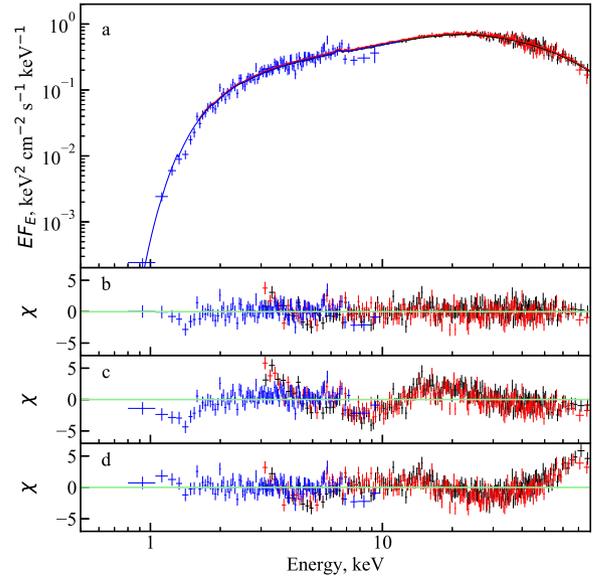}
\caption{{\it Panel a:} Unfolded spectrum of \xte\ and its approximation with the model {\sc const $\times$ tbabs $\times$ (gabs $\times$  po $\times$ highecut + gauss)} shown by a solid line. Red and black crosses show the data from the FPMA and FPMB telescopes of the {\it NuSTAR} observatory, respectively; blue crosses are for the {\it Swift}/XRT. The lower three panels show the deviations of the data from the models based on different continua: {\sc po $\times$ highecut} ({\it panel b}), {\sc cutoffpl} ({\it panel c}), {\sc comptt} ({\it panel d}). } 
\label{pic:avg_spectrum}
\end{figure}

Fig.~\ref {pic:avg_spectrum} shows the phase averaged energy spectrum of \xte\ according to the FPMA and FPMB modules of the {\it NuSTAR} observatory and the {\it Swift}/XRT telescope in the WT mode, obtained during simultaneous observations on 2015 March 31. The use of data from both observatories made it possible to cover a wide range of energies from 0.8 to 79~keV. We see that the spectrum of \xte\ has a typical shape for X-ray pulsars \citep[e.g.][]{Coburn2002,Filippova2005}.

The source spectrum is best described by an exponential cutoff power-law model ({\sc powerlaw $\times$ highecut} in {\sc xspec}). The model was modified by photoabsorption using the {\sc tbabs} component with the abundances adopted from \citet{Wilms2000}. The discontinuity at the cutoff, which resulted in artificial absorption-like residuals around the cutoff energy $E_{\rm cut}$, was ``smoothed'' using Gaussian absorption line {\sc gabs} at energy $E_{\rm gabs}$, which was tied to the parameter $E_{\rm cut}$, with the width of $\sigma_{\rm gabs}=0.1 E_{\rm cut}$~keV and the optical depth $\tau_{\rm gabs}$ \citep[see, e.g.,][for details]{Coburn2002}.  
The presence of a significant iron fluorescent K$\alpha$ line at 6.4~keV confirms its discovery by \citet{Kuhnel2016} in the {\it Swift}/XRT data. It was modeled with a Gaussian emission line {\sc gauss}.
To take into account possible systematic uncertainties in the calibration of {\it NuSTAR}/FPMA, {\it NuSTAR}/FPMB and {\it Swift}/XRT, a cross-calibration coefficient was introduced using the multiplicative component {\sc const} in the model. This constant for FPMA was fixed at 1.0 and kept free for FPMB and XRT.
The resulting best-fit parameters for the full model are summarized in Table~\ref{table:spec_params}. 

We also considered other models, usually used for the description of X-ray pulsars spectra: the exponential cutoff power-law model {\sc cutoffpl} and the Comptonized radiation model {\sc comptt} from \citet{Titarchuk1994}. But they demonstrate a significantly worse statistic and approximation quality (see Table~\ref{table:spec_params}). 

\begin{figure}
\centering
\includegraphics[width=0.8\columnwidth]{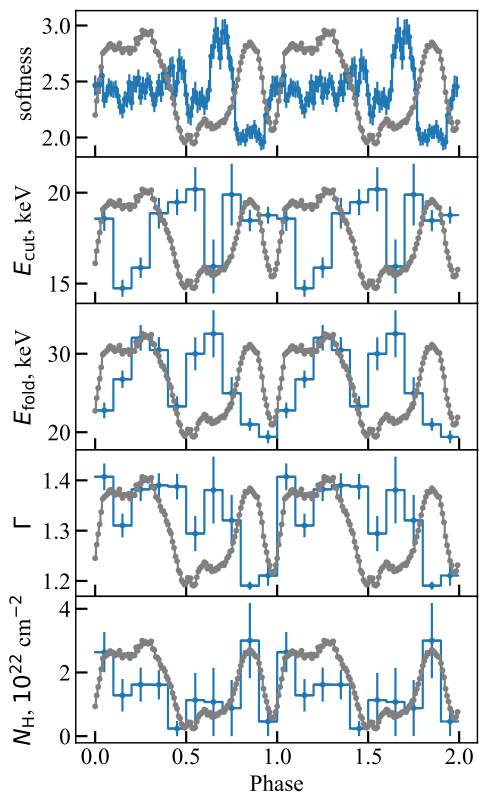}
\caption{The softness ratio of the \xte\ and the evolution of the spectral model parameters as a function of the spin phase. The averaged pulse profile in a wide energy range is superimposed in gray for visual comparison.}
\label{fig:spec_params}
\end{figure}

It is worth noting that the neutral hydrogen column density $N_{\rm H}$ shows a systematic spread depending on the continuum model used.
Nevertheless, we see in Table~\ref{table:spec_params} that regardless of the specific model, the value of $N_{\rm H}$ turns out to be 2--3 times higher than the Galactic value in the direction to the source $0.94\times10^{22}~{\rm cm}^{-2}$ obtained by \citet{HI4PI2016}.

To study the evolution of spectral parameters as a function of the rotation phase of the NS, we carried out a pulse phase-resolved spectroscopy using the {\it NuSTAR} data, divided into 10 evenly distributed phases.
To approximate the phase spectra, we used the same model {\sc const $\times$ tbabs $\times$ (gabs $\times$ po $\times$ highecut + gauss)} as for the average spectrum. The  results shown in Fig.~\ref{fig:spec_params} demonstrate significant variations in the exponential decay energy $E_{\rm fold}$, exponential cutoff energy $E_{\rm cut}$ and photon index $\Gamma$ with the pulse phase. The model used is purely phenomenological and, therefore, it is difficult to draw any physical conclusion on the origin of the evolution of spectral parameters. 
However, using spectrum softness, which is defined as the ratio of unnormalized count rate pulse profiles in the 3--10~keV / 10--20~keV energy ranges, we note the general trends in the change of the spectral form at different pulse phases (see the softness panel in Fig.~\ref{fig:spec_params}). We see that the spectrum in the interval between phases 0.6--0.8 is much softer than that in the interval 0.8--0.9.

\subsection{Optical and IR identification}
\label{sec:iden}
\begin{figure}
\centering
 \includegraphics[width=0.9\columnwidth]{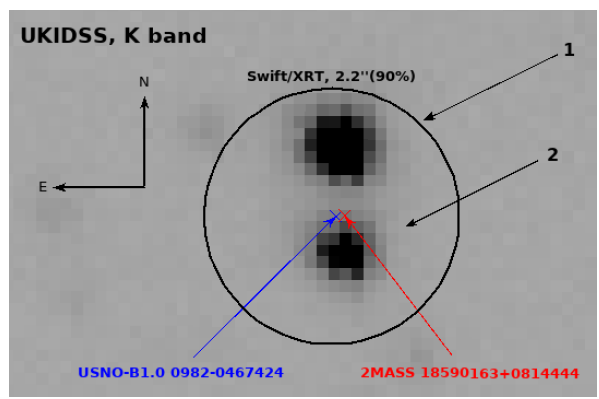}
 \caption{Image of the sky region containing \xte\ in the $K$ filter  according to the UKIDSS sky survey (UKIRT telescope). The circle shows the area of localization of the source obtained with the {\it Swift}/XRT. Possible optical/IR companions are marked with arrows and numbers. The positions of the optical companion candidate from \citet{LiKong2015} are marked with red and blue arrows.}
\label{fig:UKIDSS}
\end{figure}

In order to determine the nature of the optical companion of \xte\ we attempted to improve its X-ray localization. For that we used the XRT data collected in the PC mode during the 2015 outburst and the online tool provided by the UK \textit{Swift} Science Data Center \citep[see][]{Goad2007, Evans2009}. Unfortunately, due to a high flux from the source, all observations were affected by the pile-up effect.
Formally, according to \citet{Evans2009}, modern algorithms allow obtaining the position of the source quite accurately even in the presence of the noted effect. Nevertheless, for the further analysis we selected only one observation ObsID 00081447001 (performed on 2015 March 31, after the observation utilized by \citet{LiKong2015}) in which the pile-up effect is least pronounced. Using an additional astrometric correction with the UVOT telescope we obtained the source position: RA $=18^{\rm h} 59^{\rm m} 01\fs65$, Dec $=+08{\degr} 14\arcmin 44\farcs4 $ with the localization uncertainty of $2\farcs2$~(90\% cl), that is compatible with the result from \citet{LiKong2015}. 
A subsequent search for an optical/infrared companion of the source based on the data from the Pan-STARRS, IPHAS and UKIDSS
sky surveys revealed that two objects fall within the XRT error radius (see Fig.~\ref{fig:UKIDSS}), each of which can potentially be a companion of \xte\ (see Table~\ref{tab:XTE_IR} for their parameters). It is worth noting that the candidate considered by \citet{LiKong2015} as a potential companion of the object under study is probably the  superposition of the sources discussed in this work (see red and blue crosses in Fig.~\ref{fig:UKIDSS}).

\begin{table}
	\centering
	\caption{Possible IR/optical companions of \xte\ and their observed magnitudes.}
	\label{tab:XTE_IR}
	\begin{tabular}{lrr}
	\hline
	Number & 1 & 2 \\
	\hline
	RA & 18$^{\rm h}$59$^{\rm m}$01\fs64 & 18$^{\rm h}$59$^{\rm m}$01\fs63\\ 
	Dec & +08{\degr}14\arcmin45\farcs 4 & +08{\degr}14\arcmin43\farcs6\\ 
	$l$ & 41\fdg1348 & 41\fdg1344	\\
	$b$ &  2\fdg0767 & 2\fdg0764 \\
	\hline
	\multicolumn{3}{c}{IPHAS$_{Vega}$ ({2004 July 9})}  \\
	$r$ & $20.80\pm0.08$  & $20.76\pm0.08$\\
	H$\alpha$ &  $20.05\pm0.09$  & $20.01\pm0.10$\\
	$i$ &  $18.46\pm0.04$  & $18.79\pm0.05$\\

	\hline
	\multicolumn{3}{c}{Pan-STARRS$_{AB}$ ({2011 August 31})}  \\
	$r$ & $20.940\pm0.031$  & $20.559\pm0.027$\\
	$i$ &  $18.917\pm0.013$  & $18.972\pm0.006$\\
	\hline
	
	\multicolumn{3}{c}{UKIDSS ({2009 June 2})}  \\
	$J$ &  $14.902\pm0.003$  & $15.497\pm0.004$\\
	$H$ &  $13.792\pm0.002$  & $14.746\pm0.004$\\
	$K$ & $13.268\pm0.002$ & $14.147\pm0.005$\\
	\multicolumn{3}{c}{UKIDSS ({2012 April 25})}  \\
	$K$ & $13.250\pm0.003$ & $14.593\pm0.009$\\
	
    \hline
	\multicolumn{3}{c}{NOT ({2021 June 20})}  \\
	$Ks$ & $13.35\pm0.06$ & $13.61\pm0.06$\\

		\hline
	\end{tabular}
\end{table}

We see in Table~\ref{tab:XTE_IR} that both stars have approximately the same magnitudes in the optical filters $r$ and $i$, as well as in the H$\alpha$ filter. In the infrared range, star \#1 turns out to be somewhat brighter than star \#2 according to the archival data of the UKIDSS catalog obtained in 2009--2012. At the same time, star \#2 demonstrates a significant variability of its flux in the $K$-filter, so that during observations with the NOT telescope in June 2021, its observed magnitude was comparable to that of star \#1. Based on the above results it is not obvious which of the two optical/IR sources is the true companion of \xte. Therefore we performed dedicated spectroscopic observations in the infrared and optical bands.

\begin{figure}
\centering
\includegraphics[width=0.9\columnwidth]{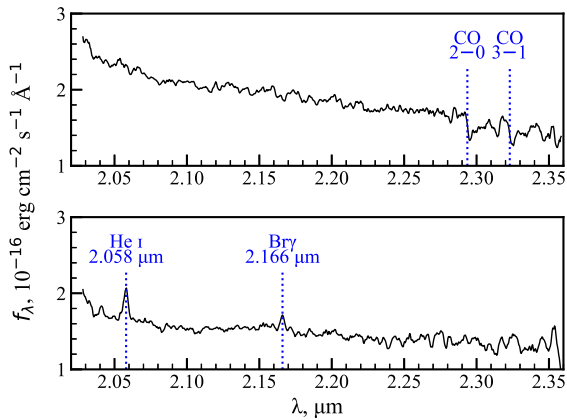}
\caption{$K$-band spectra of the two candidates for optical companion of \xte\ based on the NOT data. Upper panel is for star \#1 and the bottom one for star \#2. } 
\label{fig:NOT}
\end{figure}

First of all, using NOT facilities we performed the $K$-band spectroscopy for both candidates (see Fig.~\ref{fig:NOT}).
The spectrum of the star \#1 (the upper panel) has no sign of Br$\gamma$ emission, the spectrum is featureless apart from the presence of CO bands in absorption at 2.29 and 2.32 $\mu$m which points to a late-type star, the strength of the CO bands indicating spectral type as late as 
K or M \citep{Wallace1997}.
For star \#2 (the bottom panel), which is slightly fainter, there are clear Br$\gamma$ and \ion{He}{i} \@ 2.058 $\mu$m lines in emission, which points to a Be-type star \citep[for examples of the $K$-band spectra of Be stars, see][]{Clark2000}. 
We note that the equivalent width of \ion{He}{i} line exceeds that of Br$\gamma$. The lack of \ion{He}{ii} at
2.189~$\mu$m excludes spectral types earlier than O9~V \citep{Hanson2005}, although admittedly our spectrum has a low signal to noise and faint lines may go undetected, but it compares well to
a B1 spectral type of Be stars in the spectral atlas of \citet{Hanson1996}, and we estimate the spectral class of star \#2 as similar to B0-2Ve. Thus, based on the NOT data we consider star \#2 to be the most probable companion of the \xte, since the object behaves like a system with a Be-star in X-rays, its orbital period is consistent with possible Be-nature \citep[see][]{Corbet2009} and its X-ray outburst shape is typical for systems with Be-stars (see Fig.~\ref{fig:lightcurve}). At the same time, we cannot completely exclude that the real companion of the source is star \#1.

\begin{figure}
\centering
\includegraphics[width=0.9\columnwidth]{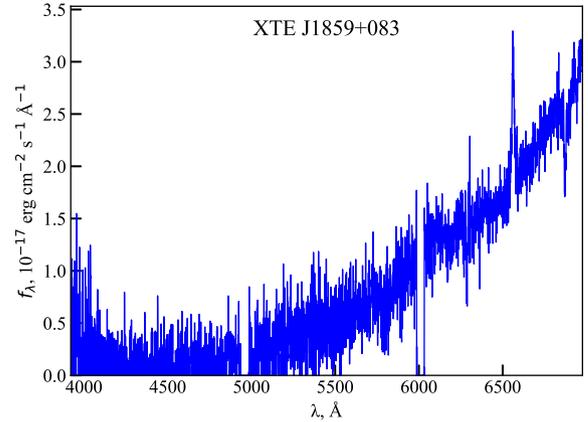}
\caption{SALT optical spectrum of the combined light of the two candidates for the optical counterpart of \xte. The two gaps are due to the CCD mosaic nature of the RSS detector.}
	\label{fig:SALT}
\end{figure}

Results from the SALT spectroscopy confirm that the optical counterpart is a Be star from the detection of a strong H$\alpha$ emission line, although it was not possible to discriminate between the two stars.  In Fig.~\ref{fig:SALT} we show the SALT optical spectrum, which is consistent with a heavily reddened Be star. The H$\alpha$ line has a FWHM of 18~\AA\ and the EW of $-$14~\AA.

According to the $Ks$-band photometry using the NOTCam acquisition images calibrated towards 2MASS stars, star \#1 has $Ks = 13.35\pm0.06$ mag, and star \#2 $Ks = 13.61\pm0.06$ mag. Interestingly, as it was already mentioned above, star \#2 turned out to be significantly brighter than in the both observations of UKIDSS survey ($K=14.147\pm0.005$ and $K=14.593\pm0.009$, see Table~\ref{tab:XTE_IR}). At the same time in the case of star \#1, the UKIDSS value ($K=13.265\pm0.002$) agrees with the estimates from the NOT observations. This indicates the possible variable nature of star \#2 in the infrared band that is often observed in Be-stars \citep[see, e.g.,][]{Dougherty1994}.  

\begin{figure*}
\centering
\includegraphics[width=0.42\textwidth]{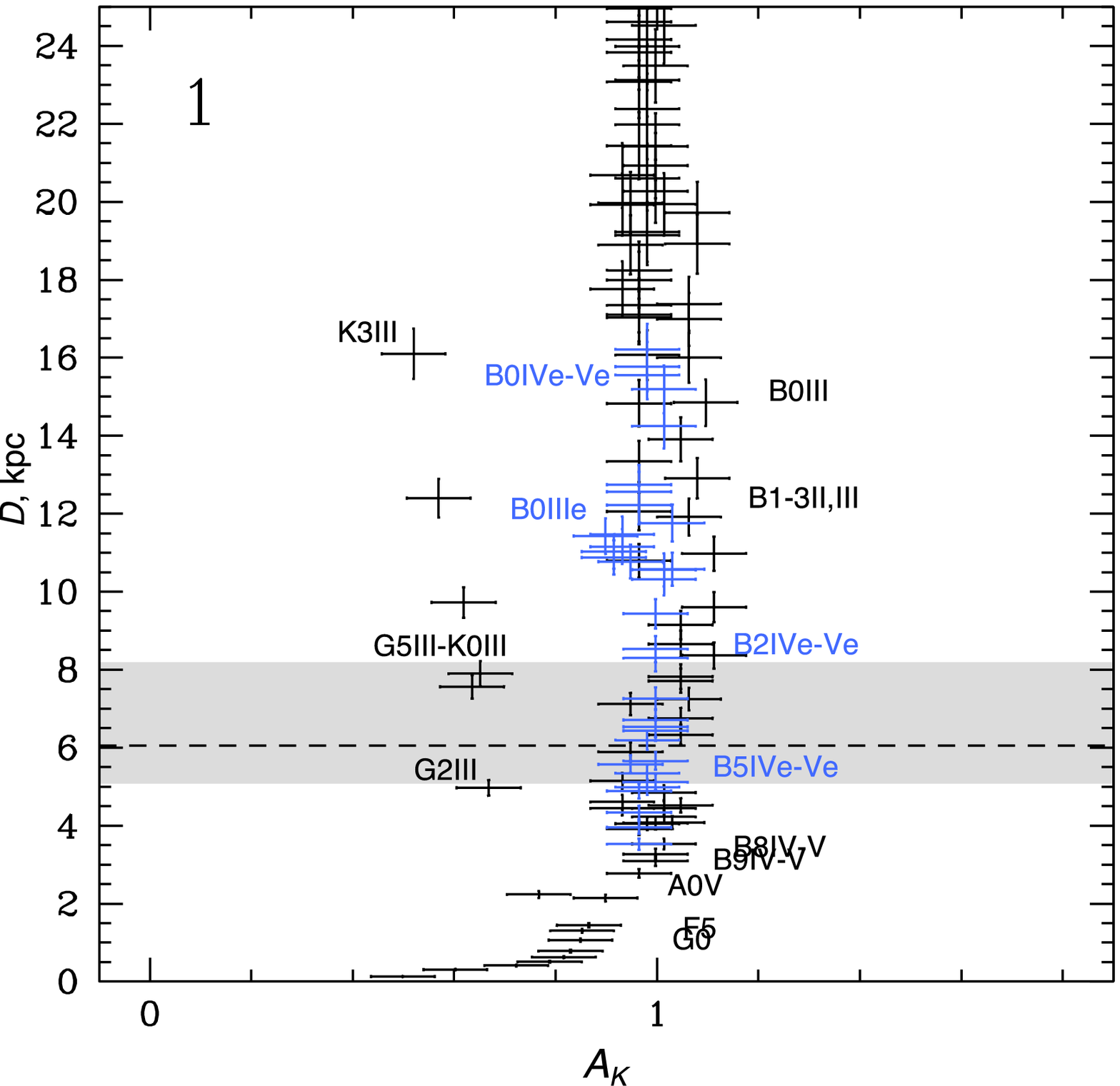}
\includegraphics[width=0.42\textwidth]{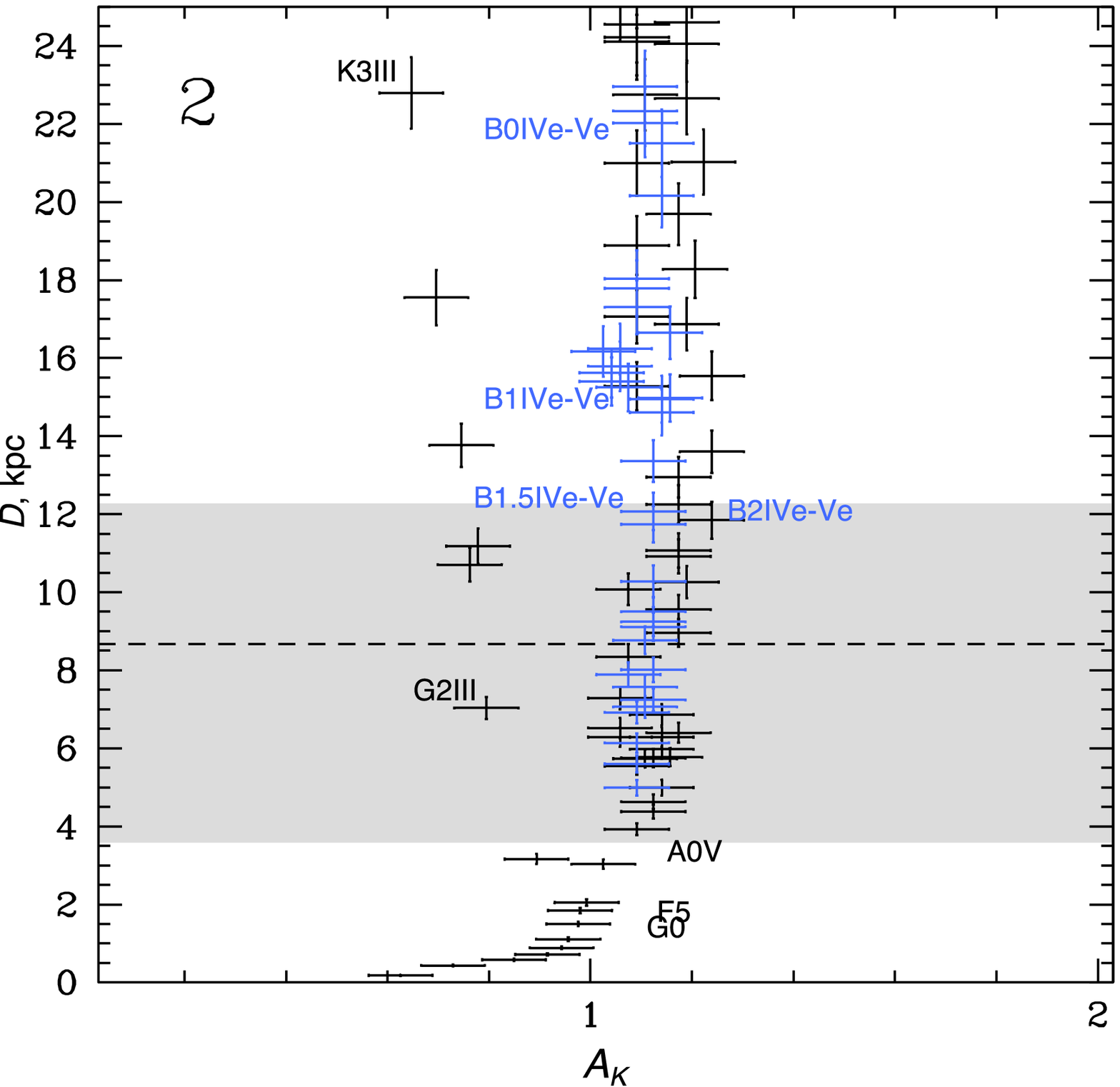}
\caption{Extinction--distance plots obtained for each of the two possible companions of \xte. They show at what distance stars of different classes should be located and how strongly they should be extincted in order to satisfy the observed magnitudes of objects in the $H$- and $K$-filters from Table~\ref{tab:XTE_IR}. Blue crosses mark the subclass of Be stars. The dashed line corresponds to the photometric distance to each of the objects taken from the work of \citet{Bailer2021}, and the shaded area corresponds to errors for this distance. 	}
\label{fig:IRTYPE}
\end{figure*}

\section{Discussion and conclusions}
\subsection{Distance estimation}

The information on the distance to the possible companions is valuable in understanding of the nature of \xte. Thanks to the {\it Gaia} observatory data, processed by \citet{Bailer2021}, we have the distance estimates for each of the putative companions. The photogeometric distance (recommended by the authors as more accurate) to star \#1 is $6.1^{+2.0}_{-1.0}$~kpc and to star \#2 is $8.7^{+3.6}_{-5.1} $~kpc. 

To check how these values compare with the estimated star classes we investigated both stars using the method described and successively applied by \citet{Karasev2015}, \citet{Nabizadeh2019}, and \citet{2021ApJ...909..154T}. Taking the absolute values of stars of various classes in the $H$ and $K$ photometric filters from \citet{Wegner2000, Wegner2006, Wegner2007, Wegner2014, Wegner2015} and applying corrections for the extinction and the distance, we determined at what distance they should be located and how strongly they should be extincted to match the observed magnitudes of the investigated companion in the $H$ and $K$ filters. The result of this approach for different possible types of the companion candidates is shown in Fig.~\ref{fig:IRTYPE}. 
We note that because the magnitude of star \#2 in the $K$ filter significantly varies with time, we provide our estimations for this object using only $H$ and $K$ magnitudes obtained on the same date on 2009 June 2. Moreover, in our analysis, we use close photometric filters $H$ and $K$, that minimize the probable effect of increasing amplitude of variations with wavelength for Be-stars \citep{Dougherty1994}.

From this approach we can estimate the extinction in the direction to the source and the distance.  Taking into account, that the star \#1 is probably a late-type star, and star \#2 is a Be-star we can estimate the extinction in their directions as $A_{K,1}\sim(0.5-0.7)$ and $A_{K,2}\sim(1-1.2)$, respectively (Fig.\,\ref{fig:IRTYPE}). Using the standard extinction law from \citet{Cardelli1989}\footnote{Because \xte\ is at a significant distance from the Galactic bulge, all estimates were made under the assumption of the standard extinction law given by \citet{Cardelli1989}, with $A_V=8.93 A_K$.} these values can be converted into the corresponding column densities of hydrogen atoms using the correlation $N_{\rm H} = 2.87\times 10^{21} A_V$ \citep{Foight2016}:  $N_{\rm H,1}\simeq (1.3-1.8)\times 10^{22}$ cm$^{-2}$ and $N_{\rm H,2} \simeq (2.6-3.1) \times 10^{22}$ cm$^{-2}$. The latter values are in good agreement with the ones measured from the source spectrum, that can be considered as an additional indication that star \#2 is the true counterpart for \xte. 

Besides, using diagram in Fig.~\ref{fig:IRTYPE}, we can also get the distance for the star \#2 as $11.7-21.5$~kpc for reasonable classes of stars (B0-2IV-Ve):
$D_{\rm B0IV-Ve}\simeq$ 21.5~kpc, $D_{\rm B1IV-Ve}\simeq$~14.6 kpc, $D_{\rm B1.5IV-Ve}\simeq$~12.1 kpc, $D_{\rm B2IV-Ve}\simeq$~11.7 kpc. 
The stars of the B1.5-2IV-Ve classes are more appropriate for star \#2 to be consistent with the {\it Gaia} distance estimations.

Additional constraints on the distance to the system can be given by using the constraint on the NS luminosity in the quiescent state. In the work of \citet{Romano2007}, the source was not detected and the authors gave a $3\sigma$  upper limit for the flux of $5 \times 10^{-14}$~\flux\, assuming a spectral model of the absorbed power-law with a photon index of 2 and the hydrogen column density of $9 \times 10^{21}\,{\rm cm}^{-2}$. We recalculated this upper limit for the source flux in the XRT observation 00037043003 using the absorbed power-law model with parameters taken from the best approximations of the broadband spectrum by the {\sc po} $\times$ {\sc highecut}  model (see Table~\ref{table:spec_params}). This gives the value of the $3\sigma$ upper limit of  $7.6 \times 10^{-14}$~\flux.
At the same time,  \citet{Tsygankov2016} and \citet{Tsygankov2017} showed that even in the absence of accretion on the NS, its luminosity, provided by the cooling of the NS crust, does not fall below $\sim10^{33}$~\lum. The lack of detection of \xte\ with the above upper limit for a given luminosity allows to constrain the distance to the system to be larger than $\sim$10~kpc, which is also in a better agreement with star \#2 as the optical companion.

\subsection{Estimation of the magnetic field strength}

One of the goals of our work was to determine the magnetic field strength of the NS in \xte. The most direct method for this is to detect the cyclotron absorption line in the energy spectrum of the source \citep[see, e.g.,][and references therein]{Staubert2019}. Our analysis did not reveal presence of such a spectral feature in the energy range 5--50~keV, which allows us to roughly limit the NS magnetic field strength to be weaker than $5 \times 10^{11}~{\rm G}$ or stronger than $5\times10^{12}~{\rm G}$. This conclusion was verified using the phase-resolved spectroscopy, as in spectra of some of X-ray pulsars the cyclotron line or its higher harmonics appear only at certain phases of rotation of the NS \citep[see, e.g.][]{2019ApJ...883L..11M,2021ApJ...915L..27M}. No significant detection of any absorption features was found using such an analysis.  

It is possible to roughly restrict the magnetic field from above by using the absence of the observed transition of the pulsar to the propeller regime \citep[see][]{IllarionovSunyaev1975} in observation {\it Swift}/XRT with the lowest unabsorbed flux of $(7.9 \pm 0.2) \times 10^{-10}$~\flux\  (ObsID 00037043014), where pulsations of the source radiation are still detected and its energy spectrum has a hard shape (see Fig.~\ref{fig:lightcurve}). Using formula~(1) from \citet{Campana2002}, we can estimate the magnetic field of the NS to be less than $2.0^{+0.9}_{-1.2} \times 10^{13}~ {\rm G}$ taking the distance to star \#2 as the most probable optical companion.

\section*{Acknowledgements}
This research was supported by the grant 14.W03.31.0021 of the Ministry of Science and Higher Education of the Russian Federation. 
The SALT observations were obtained under the SALT Large Science Programme on transients (2018-2-LSP-001; PI: DAHB) which is also supported by Poland under grant MNiSW DIR/WK/2016/07. DAHB acknowledges research support from the National Research Foundation. MG is supported by the EU Horizon 2020 research and innovation programme under grant agreement No 101004719.
This work made use of data supplied by the UK \textit{Swift} Science Data Centre at the University of Leicester and data obtained with \textit{NuSTAR} mission, a project led by Caltech, funded by NASA and managed by JPL. 
The work is partly based on observations made with the Nordic Optical Telescope, owned in collaboration by the University of Turku and Aarhus University, and operated jointly by Aarhus University, the University of Turku and the University of Oslo, representing Denmark, Finland and Norway, the University of Iceland and Stockholm University at the Observatorio del Roque de los Muchachos, La Palma, Spain, of the Instituto de Astrofisica de Canarias.
This research also has made use of the \textit{NuSTAR} Data Analysis Software ({\sc nustardas}) jointly developed by the ASI Science Data Center (ASDC, Italy) and Caltech.
\section*{Data Availability}
{\it NuSTAR} and {\it Swift} data can be accessed from corresponding online archives.
The optical and IR data underlying this article will be shared on reasonable request to the corresponding author.
 
\bibliographystyle{mnras}
\bibliography{main.bbl}

\bsp    
\label{lastpage}
\end{document}